# A low energy core-collapse supernova without a hydrogen envelope.

S. Valenti[1], A. Pastorello[1], E. Cappellaro[2], S. Benetti[2], P.A. Mazzali[2,3], J. Manteca[4], S. Taubenberger[3], N. Elias-Rosa[5], R. Ferrando[6], A. Harutyunyan[2,7], V. P. Hentunen[8,9], M. Nissinen[8], E. Pian[10], M. Turatto[11], L. Zampieri[2], S. J. Smartt[1]

1. *Astrophysics Research Centre, School of Mathematics and Physics, Queen's University Belfast, Belfast BT7 1NN, United Kingdom*

2. *INAF Osservatorio Astronomico di Padova, Vicolo dell' Osservatorio 5, I-35122 Padova, Italy*

3. *Max-Planck-Institut für Astrophysik, Karl-Schwarzschild-Str. 1, D-85741 Garching bei München, Germany*

4. *Begues Observatory, Santpere 6 Casa 22, 08859 Begues, Barcelona, Spain*

5. *Spitzer Science Center, California Institute of Technology, 1200 E. California Btvd. Pasadena, CA 91125, USA*

6. *Calle de la Guardia Civil 22, 46020, Valencia, Spain*

7. *Fundación Galileo Galilei-INAF, Telescopio Nazionale Galileo, E-38700 Santa Cruz de la Palma, Tenerife, Spain*

8. *Taurus Hill Observatory, Härcämäentie 88, 79480 Kangaslampi, Finland.*

9. *Tuorla Observatory, Department of Physics & Astronomy, University of Turku, Väisäläntie 20, FI-21500 Piikkiö, Finland*

10. *INAF Osservatorio Astronomico di Triete, Via Tiepolo 11, I-34131 Trieste, Italy.*

*11. INAF Osservatorio Astronomico di Catania, Via S. Sofia, 78, 95123, Catania, Italy*

**The final fate of massive stars depends on many factors, including mass, rotation rate, magnetic fields and metallicity. Theory suggests that some massive stars (initially greater than 25-30 solar masses) end up as Wolf-Rayet stars which are deficient in hydrogen because of mass loss through strong stellar winds. The most massive of these stars have cores which may form a black hole and theory predicts that the resulting explosion produces ejecta of low kinetic energy, a faint optical display and a small mass fraction of radioactive nickel[1,2,3]. An alternative origin for low energy supernovae is the collapse of the oxygen-neon core of a relatively low-mass star (7-9 solar masses) through electron capture[4,5]. However no weak, hydrogen deficient, core-collapse supernovae are known. Here we report that such faint, low energy core-collapse supernovae do exist, and show that SN2008ha is the faintest hydrogen poor supernova ever observed. We propose that other similar events have been observed but they have been misclassified as peculiar thermonuclear supernovae (sometimes labelled SN2002cx-like events[6]). This discovery could link these faint supernovae to some long duration gamma-ray bursts. Extremely faint, hydrogen-stripped core-collapse supernovae have been proposed to produce those long gamma-ray bursts whose afterglows do not show evidence of association with supernovae[7,8,9].**

SN2008ha was discovered on 2008 Nov 7.17 UT in the late-type galaxy UGC 12682, at an unfiltered magnitude of 18.8 (Ref. 10). A spectrum obtained on Nov. 18.18 UT (Ref. 11) was similar to that of the peculiar supernova SN2002cx at 10 days past maximum. The narrow P-Cygni lines are indicative of slowly expanding ejecta, suggesting an outflow velocity 3,000 km s$^{-1}$ slower than SN2002cx. Our spectra show narrow spectral lines confirming the low velocities of the ejected material ($v_{ej}\sim$ 2,300 km s$^{-1}$) and no signature of Hydrogen. (see Figure 1 and also SI). The spectral evolution is very fast, and [Ca II] λλ7,291-7,324 emission is detected already one month after the explosion (usually it is visible in core-collapse supernovae only after 2-3 months). The ejecta velocity slowly decreases with time, reaching ~1,500 km s$^{-1}$ in the last spectrum. Such low velocities ($v_{ej}$ ~ 1,000-1,500 km s$^{-1}$) have been observed in a group of low luminosity hydrogen-rich core-collapse supernovae[12] (faint SNe IIP) with extremely narrow P-Cygni spectral lines, but never in thermonuclear supernovae (SNe Ia). These objects are believed to be weak explosions (a few times 10$^{50}$ erg) of massive stars, ejecting only 10$^{-3}$ solar masses of radioactive material[12].

The photospheric spectrum of SN2008ha is remarkably similar to that of SN2005cs (see Figure 2), apart from the absence of hydrogen lines and of the OI λ7,774 feature, which is more pronounced in SN2008ha than in SN2005cs. The characteristic type Ia SN lines (Si II λλ6,347- 6,371 and S II λ5,640), clearly visible in SN1991T, are distinctly weak or absent in SN2008ha.

Although the similarity at early phases between SN2008ha and SN2005cs is striking, we know that photospheric spectra alone can be equivocal in discerning between thermonuclear and core-collapse explosions. The existence of prominent Si II and S II lines (combined with a lack of He I and hydrogen features) generally engenders association with the thermonuclear explosion mechanism. However, Si II lines are indeed detected in some stripped-envelope core-collapse supernovae (e.g. SN2007gr[16]). To complicate matters, the presence of nearby Fe II, Ti II and Cr II lines blended by high photospheric velocities make the identification of weak Si II rather uncertain and may cause misclassifications.

Additional constraints on the explosion mechanism can be derived from the study of late time spectra, when the ejecta are optically thin and it becomes easier to probe the nature of the innermost layers. Because of the extremely fast spectral evolution of SN2008ha, the spectrum at +65 days already shares a remarkable similarity with the nebular spectra of SN2005cs except for the hydrogen Balmer line (H$\alpha$ $\lambda$6,562 ; Figure 2b). The near-infrared Ca II triplet and emission lines due to [Ca II] $\lambda\lambda$7,291-7,324 (strong in core-collapse supernovae and absent in SNe Ia) are clearly visible in both objects. The spectrum at +65 days does not show any evidence of the prominent forbidden iron lines which dominate late-time spectra of thermonuclear supernovae. The lack of these features is a strong indication that little $^{56}$Ni was synthesised in the explosion, disfavouring a thermonuclear origin. The [O I] $\lambda\lambda$6,300-6,364 feature, which is usually prominent in stripped-envelope core-collapse supernovae, is undetected in the spectra of SN2005cs and SN2008ha, probably because of the high ejecta density at these phases.

SN2008ha is the faintest and lowest luminosity hydrogen-deficient SN known. Using a distance modulus of $\mu$=31.55 mag and a reddening of E(B-V) = 0.076 mag (see SI for details), SN2008ha has a peak magnitude of $M_R = -14.5 \pm 0.3$. This is 5 magnitudes fainter than typical type Ia SNe, and 3 magnitudes fainter than the low luminosity thermonuclear explosion. A "pseudo-bolometric" light curve is shown in Figure 3a together with light curves of the thermonuclear SN1991T and other core-collapse supernovae (SN1998bw[17], SN2007gr and SN2005cs). SN2008ha evolves much more rapidly than these other SNe and has a maximum luminosity comparable with those of low-luminosity SNe IIP[12,13].

The rapidly evolving light curve of SN2008ha, together with the modest ejecta velocities, imply very low kinetic energy and ejected mass. We roughly estimate these quantities using a toy model based on Arnett's equations[18] (see SI), obtaining an ejecta mass of 0.1-0.5 solar masses and a kinetic energy in the range 1-5 $\times$ 10$^{49}$ erg. The ejected mass is significantly smaller than the canonical 1.4 solar masses expected for thermonuclear supernovae (Ref. 19) (with the caveat that the physical values can be better constrained with more accurate modelling). The kinetic energy is also smaller than suggested by models of the pure deflagration scenario, which were proposed to explain SN2002cx-like events. We also estimate the mass of 56Ni produced in the explosion of SN2008ha to be 0.003-0.005 solar masses (see SI). This amount is very close to that observed in sub-luminous SNe IIP[12,13], but is two orders of magnitude smaller than in normal SNe Ia (0.4-0.8 solar masses of nickel, Ref. 20). We cannot exclude that some exotic thermonuclear explosion might be consistent with the observed low energy and fast light curve evolution of SN2008ha, but the observational comparisons advocate as a more likely explanation that SN2008ha was produced in the low-energy core-collapse explosion of a hydrogen-deficient massive star.

This discovery has important implications for the origin of some gamma-ray bursts. Several nearby long duration GRBs show evidence of an accompanying bright, highly energetic envelope-stripped core-collapse supernova in their light curves and spectra. Two long GRBs (GRB060614 and GRB060505, Ref. 7,8,9) were close enough that the presence of an associated supernova could be excluded down to limiting absolute magnitudes $M_R \approx$ -12.3 – -13.7. One possible explanation is that they were accompanied by an extremely sub-luminous, hydrogen poor core-collapse supernova. The discovery of SN2008ha is the first evidence that such supernovae do exist (see figure 3b).

The observed properties of SN2008ha are undeniably similar to the group of SNe similar to SN2002cx[6,15,21,22,23], and it is perhaps the most extreme object of its kind (see SI for more details on this supernova group). If SN2008ha is more plausibly explained by core-collapse then, by implication, could all SN2002cx events be explosions of this nature? So far they have been interpreted as pure thermonuclear deflagrations of 1.4 solar mass (Chandrasekhar mass) white dwarfs[24], although their observed characteristics deviate significantly from those of type Ia supernovae. Their intrinsic faintness and broad light curves are at odds with the luminosity vs. light curve shape relation (e.g.

Ref. 25) which characterises SNe Ia and itself is a consequence of comparable ejecta masses (Ref. 19). The spectra of SN2002cx-like supernovae are quite bizarre: before maximum they show similarities to luminous SNe Ia[15] (e.g. SN1991T), after maximum they are very similar to those of SN2008ha (see figure 4a) and at late time they resemble those of faint core-collapse SNe. A comparison of an unpublished late time spectrum of SN2005hk[21,22,23] (the best studied SN2002cx-like event), with that of the sub-luminous SNIIP 1997D[26] is shown in Figure 4b. The lack of forbidden lines of oxygen and iron in both spectra, and the presence of P-Cygni-type lines of FeII ~400 days after explosion (in the case of SN2005hk) is evidence of high density ($n \sim 10^8$ cm$^{-3}$, Ref. 23). Ref. 23, indeed, attempted to model a spectrum of SN2005hk at 228 days by combining a photospheric and a nebular spectrum. However the model required to reproduce the photospheric phase (a W7 model[27] scaled down to an energy of $3 \times 10^{50}$ erg) was unable to reproduce the high density of the inner ejecta at late phases. High density and low energy are sometimes found in core-collapse supernovae, particularly in faint SNe IIP[12,13].

Additional support comes from the detection of He I lines in SN2007J[28], a supernova showing remarkable similarities with SN2002cx at an earlier phase[29]. Helium lines have never been detected in thermonuclear supernovae. There is quite significant evidence for SN2008ha being a core-collapse supernova, and this family of objects could quite plausibly be of the same nature. As yet we cannot definitively rule on their explosive origins but SN2008ha is a remarkable, albeit extreme, example. Future observations of similar events will help to understand if they are indeed a form of thermonuclear explosions, low-luminosity core-collapse SNe from stripped stars of moderate mass, or the deaths of very massive stars inducing black-hole formation and fall-back.

1. MacFadyen, A. I., Woosley, S. E. & Heger, A. Supernovae, Jets, and Collapsars. *Astrophys. J.* **550**, 410-425 (2001)

**Supplementary Information** accompanies the paper on **www.nature.com/nature**.

Acknowledgements and the Competing Interests statement:

This work, conducted as part of the European Science Foundation EURYI Awards scheme, was supported by funds from the Participating Organisations of EURYI and the EC Sixth Framework Programme. The work of SB, EC and MT was supported by grants of the PRIN of Italian Ministry of University and Sci. Reaserch. This paper is based on observations collected at TNG, NOT, LT (La Palma Canary Island, Spain), at Ekar (Asiago Observatory, Italy), at the Begues Observatory and Arguines Observatory telescopes (Barcelona and Segorbe, Spain), at the Taurus Hill Observatory (Varkaus, Finland), at the Calar Alto Observatory (Spain) and at the ESO-UT2 (Paranal, Chile). Our analysis included data from SUSPECT Archive (http://bruford.nhn.ou.edu/~suspect/index1.html).. This manuscript made use of information contained in the Bright Supernova web pages (D. Bishop), as part of the Rochester Academy of Sciences.

Correspondence and requests for materials should be addressed to Stefano Valenti (s.valenti@qub.ac.uk).

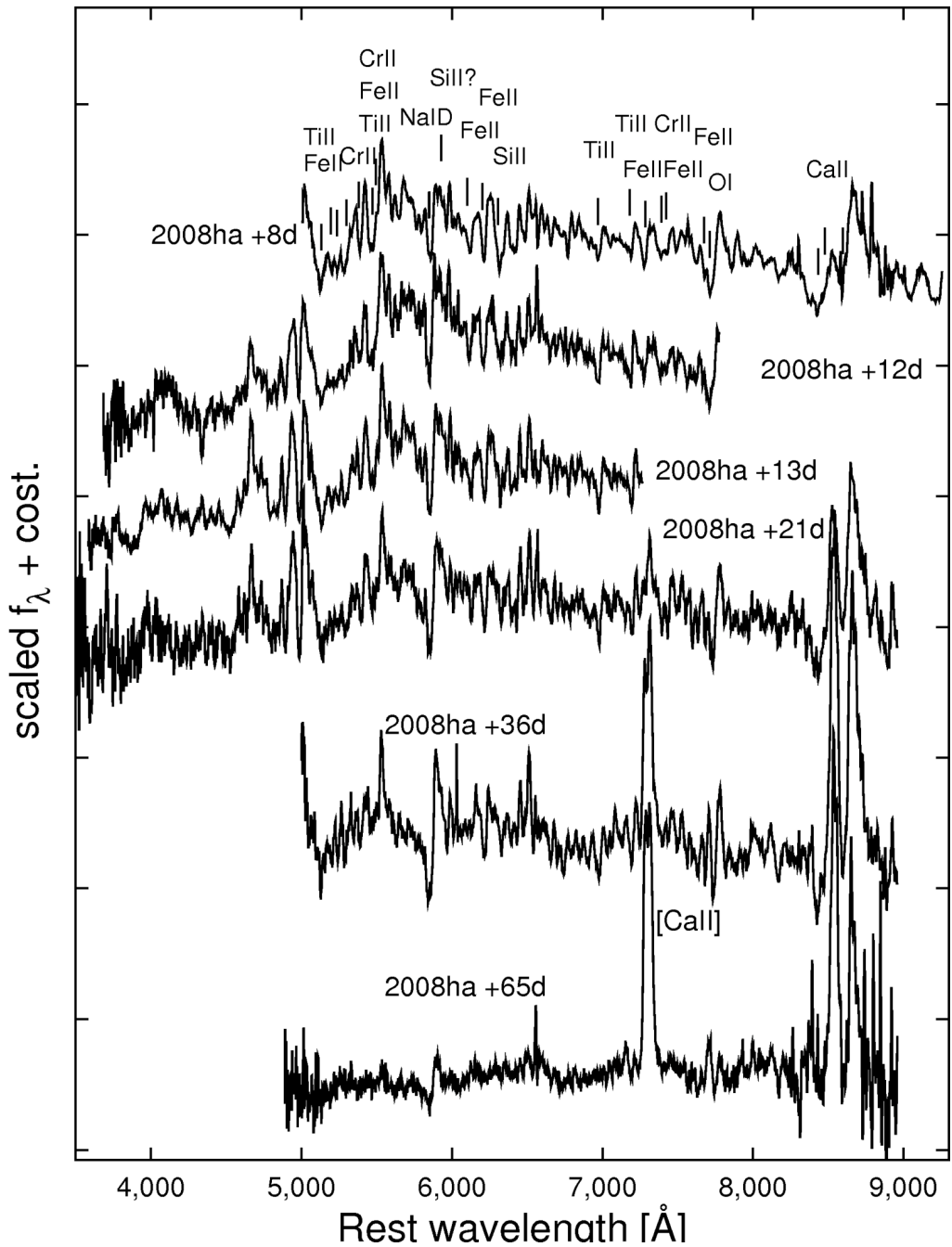

**Figure 1 | Spectral evolution of SN2008ha.** Spectral evolution of SN2008ha from +8 to +65 days after R-band maximum (which we estimate to have occurred on JD = 2454787±2). There is no signature of hydrogen features in the spectra, while Fe II, Ti II, Cr II, Na ID, O I and Ca II are clearly detected. The expansion velocity (measured from the minimum of the Fe II lines) is slowly decreasing from 2,300 to 1,500 km s$^{-1}$ in the

covered spectral range suggesting a very slow velocity evolution. The low velocity of SN2008ha allows us to identify most lines which are normally blended. Fe II and other metal lines are identified, and those of intermediate-mass elements (Na I, Ca II, O I) are prominent at all epochs (see Supplementary Information, Figure 3). On +36 days strong lines of [CaII], typical of core-collapse supernovae, become visible, confirming the fast evolution to the nebular phase of SN2008ha. The low expansion velocity and the fast spectral evolution suggest an extremely low kinetic energy and ejected mass of SN2008ha (1-5 x $10^{49}$ erg and 0.1-0.5 solar masses respectively). These values are inconsistent with a thermonuclear scenario and quite common in faint core-collapse supernovae.

**Figure 2 | Comparison of spectra of SN2008ha with those of other SNe. (a)** The spectrum of SN2008ha taken 8 days after maximum is compared with that of the under-luminous type IIP SN2005cs[13] during the hydrogen recombination, with those of the luminous thermonuclear SN1991T[14] (this sub-type of SNe Ia have been proposed to share some spectral similarities with SN2002cx[15]) and the type Ic SN2007gr[16] at comparable phases. The spectra of SN2008ha and SN2005cs are very similar, except for the H lines (always prominent in SNe IIP) and the OI feature at $\lambda 7,774$. The photospheric spectra of type Ia and type Ic supernovae both share some similarities with SN2008ha only if the spectra are red-shifted (by an ad-hoc velocity) or the spectra of SN2008ha being smoothed with a Gaussian filter. **(b)** The latest spectrum of SN2008ha at 65 days past the R-band maximum has not fully completed the transition to the nebular phase. However the remarkably fast evolution of this event means that it can be compared with the nebular spectra of other SNe. SN2008ha is very similar to the nebular spectrum of SN2005cs[13] except for the distinct lack of hydrogen features. The prominent [Fe II] lines typical of thermonuclear supernovae and the [OI] lines of stripped-envelope core-collapse supernovae are not apparent in SN2008ha.

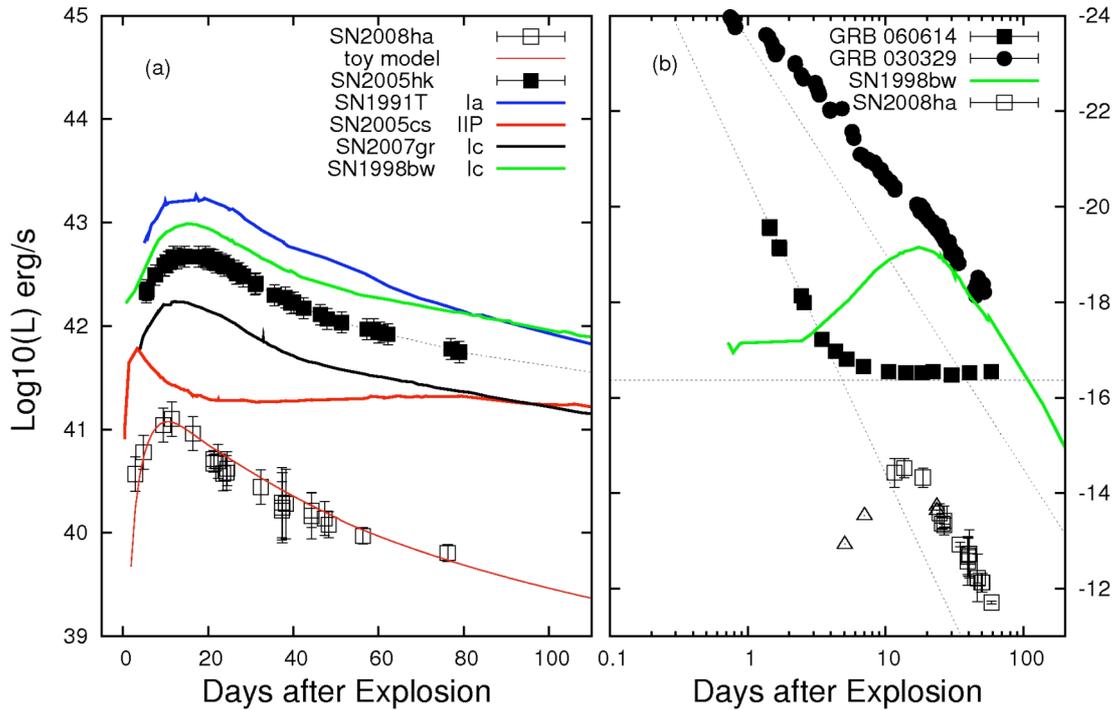

**Figure 3 | Pseudo-bolometric and absolute R-band light curves of SN2008ha. (a)** We compare a pseudo-bolometric light curve of SN2008ha with those of other SN types (error bars show the standard errors). The pseudo-bolometric light curve of SN2008ha was computed, using our R-band observations and SN2005hk as references (after having time stretched the data of SN2005hk to fit the time evolution of SN2008ha). The light curve of SN2008ha is faster than those of most other supernovae. Considering the low expansion velocity, the light curve of SN2008ha is inconsistent with a thermonuclear explosion of 1.4 solar masses typical of SNe Ia. The peak luminosity is similar to SN2005cs, suggesting possibly a stripped core analogue of this type of explosion[13], or the weak explosion of a very massive star with black hole formation[1,2,3] **(b)** The R-band light curve of SN2008ha is shown together with those of SN1998bw[17] and the afterglows of two long-duration GRBs (error bars show the standard errors). The detection of Ref. 10 and magnitudes from the Bright Supernova web site (http://www.rochesterastronomy.org/snimages/ ) are reported, with triangles.

GRB060614 showed no evidence for a SN signature in its R-band afterglow light curve leading Ref. 7,8,9 to suggest that a faint supernova (fainter than -13.7 at peak) would be required to be consistent with the canonical physical production mechanism for long GRBs. The horizontal line is the host galaxy of GRB060614. The light curve of SN2008ha shows, for the first time, that such faint hydrogen poor core-collapse supernovae (even though SN2008ha was slightly brighter than -13.7 at maximum (being R=-14.5±0.3), do exist. As a comparison, the afterglow of a bright GRB (GRB030329) consistent with the explosion of a SN1998bw-like event is also shown[30]. In that case the flux excess with respect to the afterglow (dotted line) was partially due to the SN1998bw-event.

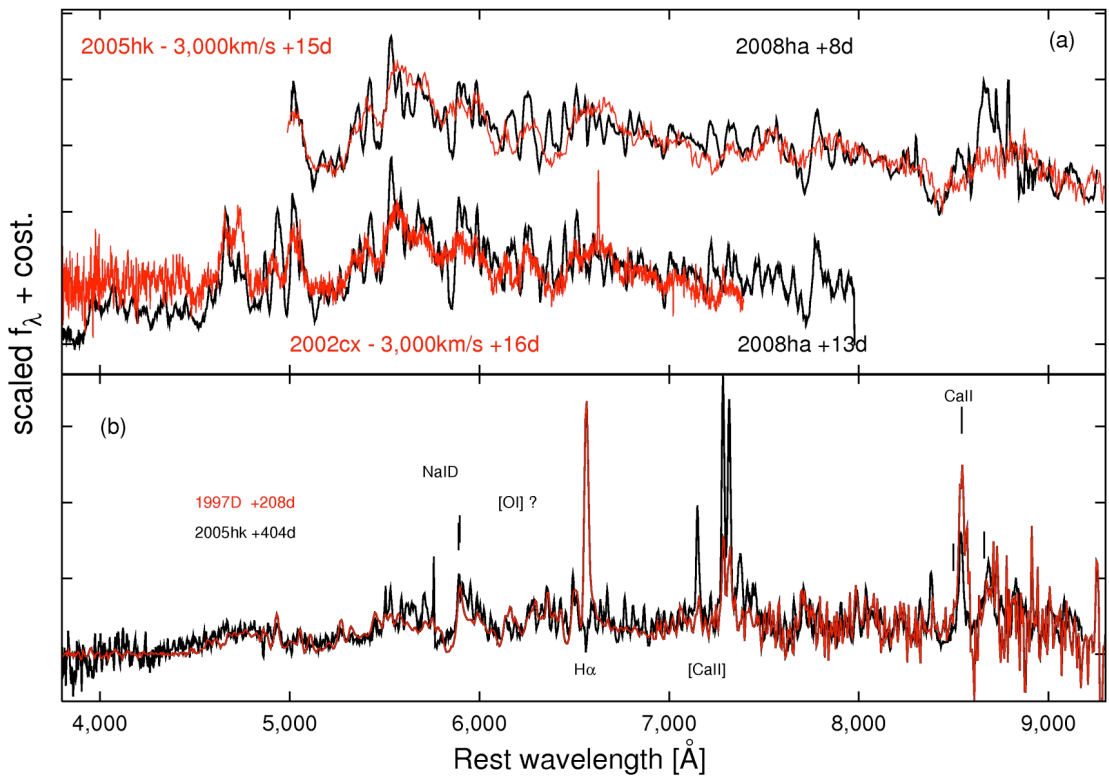

**Figure 4 | Spectra of SN 2008ha and SN2002cx like events. (a)** The spectrum of SN2008ha is compared with those of SN2002cx[6,15,21] and SN2005hk[22,23] at 2-3 weeks after maximum light. The most significant differences between the spectra are due to a different degree of line blending. The spectra of SNe 2005hk and 2002cx are shifted by -3,000 km s$^{-1}$ to align the absorptions. **(b)** The nebular spectrum of SN2005hk at ~400 days is compared with that of the faint hydrogen-rich supernova SN1997D[26]. In analogy to SN1997D, the nebular spectrum of SN2005hk is rich in permitted Fe II lines, clear evidence that the ejecta are still dense and not completely transparent. The [Fe II] and [Fe III] lines typical of thermonuclear supernovae, and the [O I] typical of stripped-envelope core-collapse supernovae are not detected. In SN1997D the [O I] feature appeared only about 6 months later[26]. The spectrum of SN2005hk was taken on 2006 November 27$^{th}$ at VLT+FORS2 (prog. ID 078.D0246) (Stanishev et al. in prep.).

# Supplementary Information for "A low energy core-collapse supernova without a hydrogen envelope." (Valenti et al.)

## 1. Optical observations

SN2008ha was monitored in photometry and spectroscopy using the following telescopes: the 3.58-m Telescopio Nazionale Galileo (TNG) equipped with Dolores, the 2.5-m Nordic Optical Telescope (NOT) with ALFOSC, and the 2.0-m Liverpool Telescope (LT) with RATCam, all located at the Roque de los Muchachos in La Palma (Canary Island, Spain). Additional data have been collected at the 1.82-m Copernico Telescope of the Asiago Observatory (Mt. Ekar, Asiago, Italy) , the 0.36-m Schmidt Telescope of the Begues Observatory (BO) in Barcelona (Spain), the 0.40m Schmidt Telescope of the Arguines Observatory (AO) in Segorbe (Spain), the 2.2m Telescope at the Calar Alto Observatory (Spain), and the 0.30-m Telescope of the Taurus Hill Observatory (THO) in Varkaus (Finland). Supplementary Table 1 reports the log of observations spanning two months after the discovery. In columns 1 and 2 the date and the Julian Day of the observations are given, in column 3 and 4 the telescope name and the set-up (respectively) used in the observations are reported, while in column 5 and 6 the V and R band magnitudes measured with the PSF-fitting technique and calibrated using Landolt fields[31]. Those were observed during two photometric nights (marked in the table with *). SN2008ha reached its maximum in the R band (17.2 ±0.3) on JD=2454787±2. Since there is no spectroscopic evidence of interstellar Na ID lines at the SN rest frame, we assume the host-galaxy extinction to be negligible. Hence, we applied reddening correction only for the Milky Way extinction, viz. E(B-V)= 0.076 (Ref. 32). Adopting a distance modulus of $\mu$=31.55 mag (see Supplementary Table 2), this gives an absolute R-band magnitude at peak $M_R$= -14.5. Starting from the R light curve, we computed a pseudo-bolometric light curve.

Since the spectral energy distribution of SN2008ha is very similarity to that of SN2005hk, we used this object as a reference and add a bolometric correction to the R-band flux (taking also in to account the different time evolution of the two supernovae). Fitting the pseudo-bolometric light curve with a toy-model based on Arnett's equations (Ref. 18) (see Ref. 33 for details on the toy-model) we estimated for SN2008ha an ejected mass between 0.1 and 0.5 solar masses, a kinetic energy of 1-5 x $10^{49}$ erg, a nickel mass of 0.003-0.005 solar masses and an JD ~ 2454773 as explosion time. Even though these values should be confirmed by detailed models, such a small ejected mass is consistent with the fast luminosity and spectral evolution of SN2008ha, and makes the thermonuclear explosion of a degenerate white dwarf of 1.4 solar masses (i.e. a normal SNe Ia), very unlikely.

## Supplementary Table 1: Photometric and spectroscopic observations of SN2008ha.

| Date | JD | Telescope | Setup | V | R |
|---|---|---|---|---|---|
| **Photometry** | | | | | |
| 2008-11-13 | 2454784.26 | BO | Unfiltered | -- | 17.30 ± 0.30 |
| 2008-11-16 | 2454786.29 | BO | Unfiltered | -- | 17.20 ± 0.20 |
| 2008-11-21 | 2454791.29 | BO | Unfiltered | -- | 17.40 ± 0.20 |
| 2008-11-26 | 2454797.29 | BO | Unfiltered | -- | 18.15 ± 0.20 |
| 2008-11-27 | 2454798.46 | Ekar | AFOSC-BVRI | 18.86 ± 0.15 | 18.36 ± 0.12 |
| 2008-11-28 | 2454799.24 | THO | Unfiltered | -- | 18.30 ± 0.30 |
| 2008-11-28 | 2454799.37 | TNG | DOLORES-Unfiltered | -- | 18.20 ± 0.05 |
| 2008-12-06 | 2454807.37 | NOT | ALFOSC-UBVRI | 19.25 ± 0.09 | 18.80 ± 0.05 |
| 2008-12-11 | 2454812.27 | BO | Unfiltered | -- | 19.03 ± 0.40 |
| 2008-12-11 | 2454812.32 | AO | Unfiltered | -- | 19.15 ±0.40 |
| 2008-12-11 | 2454812.36 | LT | RATCAM+VR | 19.54 ± 0.10 | 19.06 ±0.30 |
| 2008-12-12 | 2454813.27 | BO | Unfiltered | -- | 19.00 ± 0.40 |
| 2008-12-18 | 2454819.23 | AO | Unfiltered | -- | 19.50 ± 0.40 |
| 2008-12-18 | 2454819.25 | LT | RATCAM+VR | 19.94 ± 0.15 | 19.50 ± 0.40 |
| 2008-12-21* | 2454822.39 | TNG | DOLORES-VR | 19.99 ± 0.05 | 19.60 ± 0.20 |
| 2008-12-22 | 2454823.38 | LT | RATCAM+VR | 20.15 ±0.40 | 19.60 ±0.20 |
| 2008-12-27 | 2454828.25 | THO | Unfiltered | -- | < 19.5 |
| 2008-12-30* | 2454831.39 | CA | CAFOSC+BVRI | 20.37 ±0.13 | 20.02 ± 0.10 |
| 2009-01-01 | 2454833.18 | THO | Unfiltered | -- | <20.0 |
| 2009-01-19 | 2454851.38 | TNG | DOLORES-VR | 20.82 ± 0.4 | 20.49± 0.20 |
| **Spectroscopy** | | | | | |
| 2008-11-23 | 2454794.38 | TNG | DOLORES-LRR | | |
| 2008-11-27 | 2454798.39 | Ekar | AFOSC-GR4 | | |
| 2008-11-28 | 2454799.38 | TNG | DOLORES-LRB | | |
| 2008-12-06 | 2454807.42 | NOT | ALFOSC-gm4 | | |
| 2008-12-21 | 2454822.40 | TNG | DOLORES-LRB/R | | |
| 2009-01-19 | 2454851.36 | TNG | DOLORES-LRR | | |

## 2. The SN2002cx-like family

We found, in the literature, 7 supernovae which have been reported to be similar to SN2002cx (Refs. 6,15,21,22,23,34,28,35,36,37) and these are listed in Supplementary Table 2. One of them, SN2007J, was later re-classified as a peculiar helium-rich (SNIb) core-collapse SN, since He I features appeared later in the spectra[28] (see main manuscript).

SN2007qd (another SN of this class) was quite similar to SN2008ha. It showed a photospheric velocity of ~2,600 km s$^{-1}$ and was 1 magnitude fainter than SN2002cx (Ref. 34). Ref. 23 suggested, for SN2007qd, in the hypothesis of a thermonuclear explosion of 1.4 solar masses of ejecta, a W7 model[27] scaled down to an energy ~ 8 x10$^{49}$ erg, which would result in a very slow evolution of the light curve. This confirms that the fast evolution of the light curve of SN2008ha is in disagreement with a thermonuclear explosion of 1.4 solar masses.

We also investigated the galaxy types of SN2002cx-like events (see Supplementary Table 2). The galaxy types have been collected from the NED (Ref. 38) and HyperLeda (Ref. 39) databases, with the exception of the hosts of SN2002cx and SN2007qd. Images of these 2 galaxies, from the SDSS DR7 (Ref. 40), are shown in Figure 1, and we estimate for them morphological types Sb/Sc and SBb/SBc respectively. The distance for SDSS J020932.74-005959.6 was taken from SDSS DR7. The galaxy absolute magnitude values $M_B$ and $M_g$ are from Hyperleda and SDSS DR7. The host galaxies are not particularly faint, suggesting that there is no evidence for dwarf hosts or low metallicity environments for the progenitor stars. In Figure 2, we compare the host galaxies morphology types of SN2002cx-like events with those of other SN types. All eight SNe have late-type host galaxies (Sb or later) (see Figure 4, and Table 2 of Supplementary Information). After the submission of our work, another paper on SN2008ha and SN2002cx-like events was submitted (Ref. 41). They consider a larger sample of 15 candidate SN2002cx-like events and find that one is hosted in an S0 galaxy (SN2008ge in NGC1527) and two others are hosted in Sa type spirals (SN2006hn and SN2008A). Remarkably, the galaxy morphology distribution of SN2002cx-like events resemble that of the star formation distribution of Ref. 42, with signature of star formation also in S0.

In our sample and in the sample of Ref 41, there is some evidence that the host galaxy distribution of SN2002cx-like objects is different from that of normal and under luminous thermonuclear explosion and is consistent with either all other CC SN classes and high-luminous 1991T-like thermonuclear SNe. However, the statistics in both our sample and the extended sample of Ref 41 are somewhat too small to draw firm conclusions from numerical statistical tests. The most we can say is that there is does appear to be a lack of SN2002cx-like objects in early-type galaxies and their distribution is not unlike core-collapse SNe.

**Supplementary Table 2: Host galaxies of SN2002cx-like events.**

| SN | Host Galaxy | Type | $M_\lambda$ | Distance Modulo | Source |
|---|---|---|---|---|---|
| 2002cx | CGCG 044-035 | Sb/Sc | −19.7 (B) | 35.02 | HST archive image |
| 2003gq | NGC7407 | Sbc/Sc | −21.70 (B) | 34.80 | NED/Hyperleda |
| 2005P | NGC5468 | SAB(rs)cd/SABc | −20.3 (B) | 33.01 | NED/Hyperleda |
| 2005cc | NGC5383 | SBb (pec Sbrst)/Sb | −20.71 (B) | 32.65 | NED/Hyperleda |
| 2005hk | UGC272 | S(AB)(s)d/SABc | −19.42 (B) | 33.65 | NED/Hyperleda |
| 2007J | UGC1778 | SAdm/Sd | −20.24 (B) | 34.27 | NED/Hyperleda |
| 2007qd | SDSS J020932.74 | SBb or SBc | −19.7 (g) | 36.12 | SDSS |
| 2008ha | UGC12682 | Im/Irr | −18.14 (B) | 31.55 | NED/HyperLEDA |

**Supplementary Figure 1| Host galaxies of SN2007qd and SN2002cx.**

The image of the host galaxy of SN2007qd SDSSJ020932.74-005959.6 is shown on the left, with the position of the SN marked by cross-hairs. The host is a spiral at $z = 0.043$, with a bar oriented east-west (SBb or SBc). The host galaxy of SN2002cx (CGCG 044-035), is shown on the right. A high resolution HST WFPC2 archive image of CGCG 044-035 clearly shows it to be an inclined Sb/Sc galaxy with well defined spiral structure.

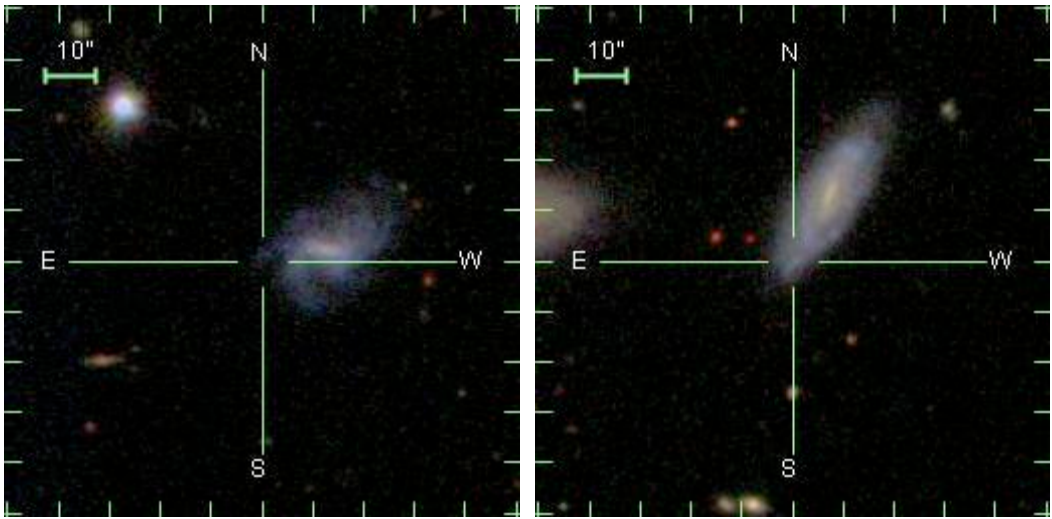

**Supplementary Figure 2| Host galaxies of SN2002cx-like events.** The distribution of the host-galaxy morphology of 2002cx-like events (red) is compared with those of under luminous and normal thermonuclear supernovae, H-rich core-collapse supernovae and H-poor core-collapse supernovae (blue). Information on the host galaxies of 2002cx-like events is reported in Table 2 of Supplementary Information, while the data of the hosts of other supernovae are from the Asiago catalogue (http://web.oapd.inaf.it/supern/cat/). SN2002cx-like events occur mainly in late-type galaxies, (Sb or later), while under-luminous and normal SNe Ia occur also in early-type galaxies. In particular, under-luminous SNe Ia, those that share with SN2002cx-like events some similarity in the peak luminosity, occur mainly in S0 or earlier.

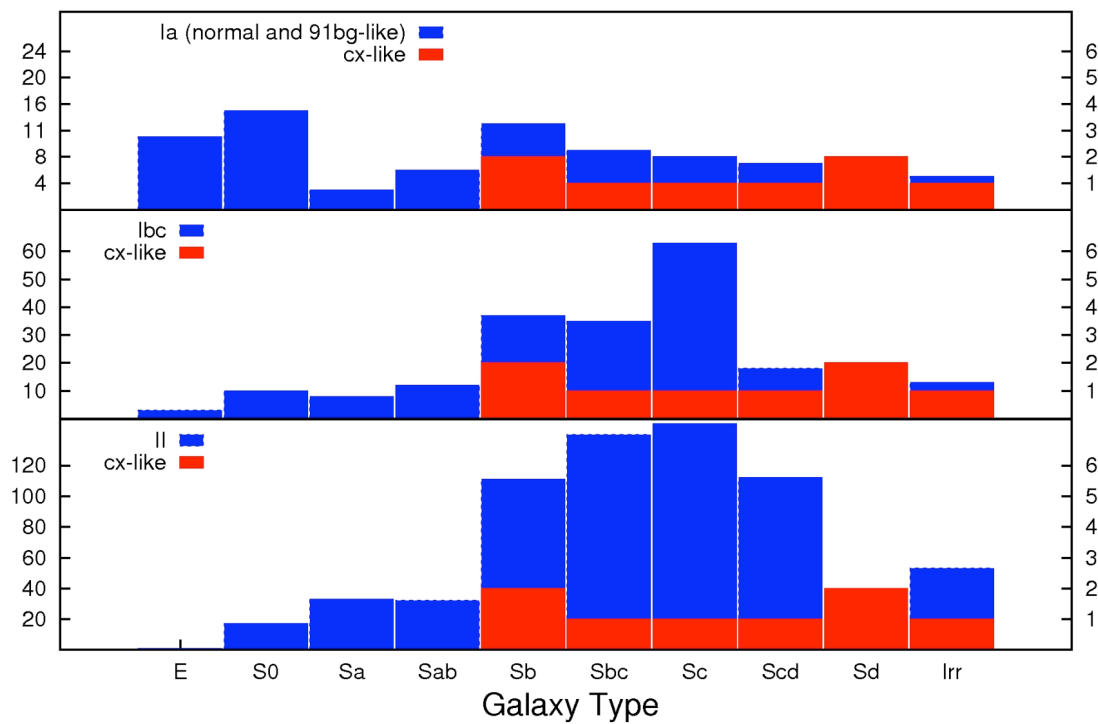

**Supplementary Figure 3| Spectroscopic comparison between SN2008ha and SN2007gr.** The spectra of SN2008ha are compared with those of the type Ic SN2007gr at similar epochs. The spectra of SN2007gr are shifted to longer wavelengths by 3000 km s$^{-1}$ in order to account for the different photospheric velocities. All major features of SN2007gr appear also in the spectra of SN2008ha, but with narrower profiles. This comparison shows that, during the photospheric phase, the intermediate-mass elements, identified in SN2008ha, are common also in other core-collapse SNe.

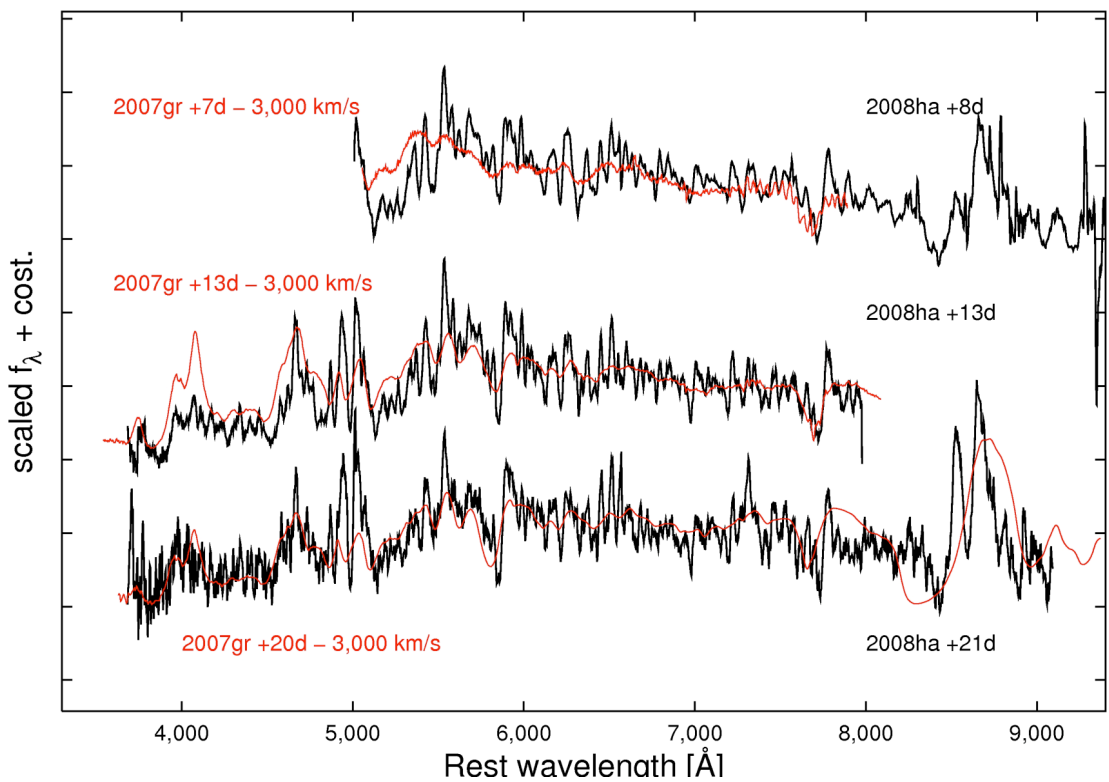

**Supplementary Figure 4| Line identification in SN2008ha.** The narrow spectral lines give us the possibility to identify the ions in the early spectrum of SN2008ha. Using Synow (Ref. 43), a spectrum synthesis code particularly suited for line identification, we obtain a good fit to the SN2008ha spectrum at 13 days after maximum using the following parameters: a continuum blackbody temperature of $T_{bb}$ = 6,200° K, a photospheric velocity v=1,300 km s$^{-1}$, a power-law index for the radial density profile of n = 6 and an excitation temperature of individual ions in the range 5,000-6,000° K for neutral ions and 8,000-10,000 for single ionized species. We were able to reproduce the observed spectrum reasonably well using 10 species (Fe II, Ti II, Cr II, O I, He I, Si II, Sc II, Mg II, Na I, Ca II). Fe II, Ti II and Cr II produce most of the observed lines in the spectrum, while other ions were included to improve the fit. Vice versa, including S II did not improve the fit. Detailed spectral modelling is needed to confirm the line identification and derive element abundances.

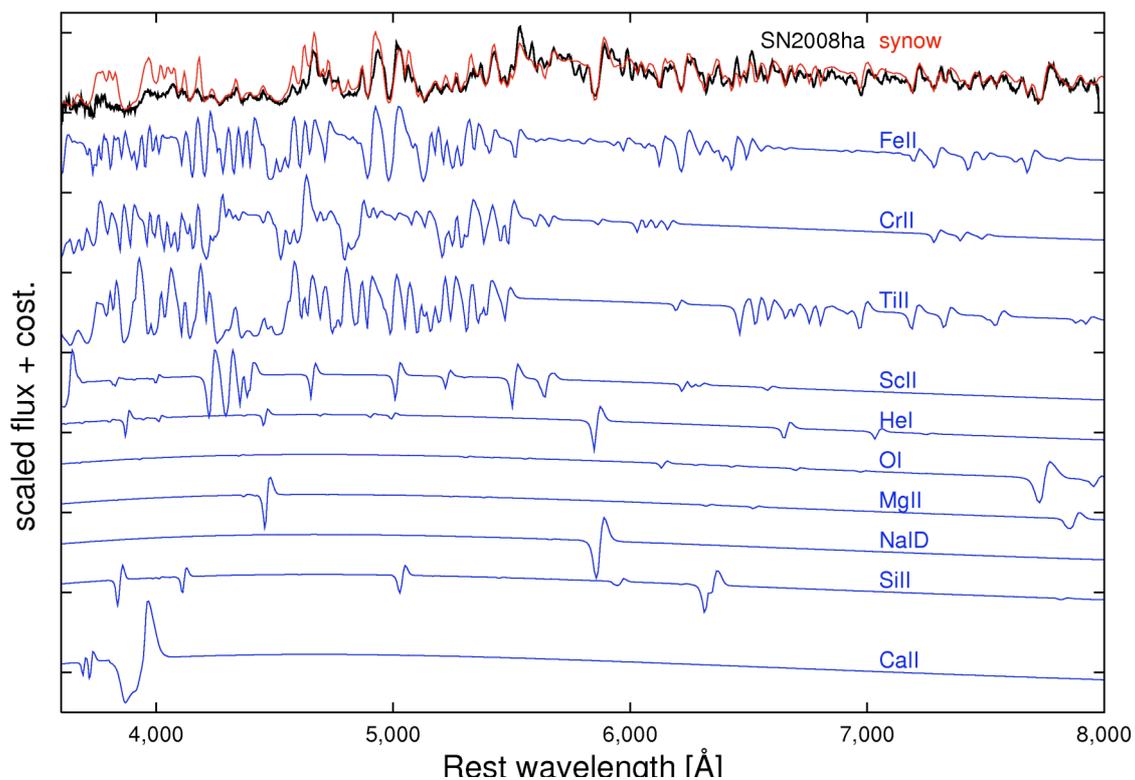